\documentclass{emulateapj}

\newcommand{\etal}{{\it et al.}}

\newcommand{\be}{\begin{equation}}
\newcommand{\ee}{\end{equation}}

\begin{document}
\title{Morphology, Environment, and the HI Mass Function}

\author{Christopher M. Springob\altaffilmark{1}, Martha P. Haynes\altaffilmark{1,2}, and Riccardo Giovanelli\altaffilmark{1,2}}

\altaffiltext{1}{Center for Radiophysics and Space Research, Cornell University, Space Sciences Building, Ithaca, NY 14853; springob@astro.cornell.edu, haynes@astro.cornell.edu, riccardo@astro.cornell.edu}
\altaffiltext{2}{National Astronomy and Ionosphere Center, Cornell University, Ithaca, NY 14853.  The National Astronomy and Ionosphere Center is operated by Cornell University under a cooperative agreement with the National Science Foundation.}

\hsize 6.5 truein
\begin{abstract}
We exploit a large, complete optical diameter and HI flux limited sample of spiral galaxies with types later than S0a to derive a robust measurement of the HI mass function (HIMF) for masses log$(M_{HI}/M_{\odot}) > 7.4$ which takes into account the effects of local large scale structure.  The global HIMF derived for this optically--selected sample is well fit by a Schechter function with $\alpha = -1.24$, log$(M_{*}/M_{\odot})=9.99$, $\phi_{*}=3.2 \times 10^{-3}$ Mpc$^{-3}$. These values match those derived from blind HI surveys to within the estimated uncertainties, yet our estimated HIMF is clearly lower than most other estimates at the lowest masses.  We also investigate the variation in the derived HIMF among spiral subclasses, finding a clear distinction between the Schechter parameters found for types Sa-Sc and those Scd and later, in the sense that the HIMF of the latest types is rising at the low mass end, whereas that of the  main spiral classes is flat or even declining.  We also explore the possible environmental dependence of the HIMF by computing it separately in regimes of differing cosmic density. The HIMFs of higher density regions are found to have flatter low-mass ends and lower values of $M_{*}$ than those of lower density regions, although the statistical significance of the difference is low.  Because the subsamples found in different density regimes exhibit virtually the same morphological fractions, the environmental dependence cannot be accounted for by morphological segregation, and must be a consequence of differences among galaxies of the same morphological type but found in different environments.  If this dependence is caused by the well known deficiency of galaxies in clusters, then it would suggest that galaxies of small linear optical diameter are characterized by higher HI deficiency, an expectation consistent with gas removal mechanisms such as ram pressure stripping.

\end{abstract}


\section {Introduction}

Because of its abundance in the interstellar medium and its emission strength at centimeter wavelengths, the neutral hydrogen content of galaxies serves as a good probe of galaxy environment and as an indicator of the potential for future star formation.  For this reason, the probability distribution over HI mass, the HI mass function (HIMF), can provide insight into the processes of galaxy formation and evolution.  It is well known that a significant fraction of cluster galaxies are deficient in HI relative to field galaxies (see, e.g., Haynes, Giovanelli, \& Chincarini 1984).  Yet, to date, studies of the environmental dependence of the HIMF have been limited to comparisons of the HIMF for galaxies in or near a particular cluster or void and those in the field (Briggs \& Rao 1993; Szomoru \etal ~1996; Verheijen \etal ~2001; Rosenberg \& Schneider 2002, hereafter RS; Davies \etal ~2004; Gavazzi \etal ~2004).  A more comprehensive study of the link between the environmental dependence of the HIMF and the deficiency of HI in clusters may shed light on the physical processes responsible for the gas depletion.

While most early studies of the HIMF relied on optically--selected samples (e.g., Briggs \& Rao 1993; Solanes, Giovanelli, \& Haynes 1996, hereafter SGH), attention has shifted more recently to results obtained from ``blind'' HI surveys (e.g., Zwaan \etal ~1997; RS; Zwaan \etal ~2003, hereafter Z03).  Unlike optically--selected samples, HI-selected samples have no bias against the low luminosity and low surface brightness galaxies which may be underrepresented in optical catalogs.  However, as noted by SGH, HI blind surveys include their own selection effects.  Notably, HI selected samples to date have tended to sample a much shallower volume than optical surveys, potentially leading to systematic errors in HI mass estimates due to distance uncertainties (Masters, Haynes, \& Giovanelli 2004), and resulting in the selection of a population of objects that is not representative of the population at larger distances.

The largest HI blind survey to date is the HI Parkes All Sky Survey (HIPASS; Meyer \etal ~2004, Zwaan \etal ~2004).  Z03 derived an HIMF from the HIPASS Bright Galaxy Catalog (BGC; Koribalski \etal ~2004).  Those authors also cross-correlated the BGC with known optical galaxies, and identified previously-catalogued optical counterparts for 892 of the 1000 HI detections.  As also found by RS, the HI blind samples include significant proportions of late-type low optical surface brightness systems that are not found in optical apparent magnitude or diameter limited catalogs but are not otherwise unusual objects. The large fraction of BGC detections found to have previously-known optical counterparts was used to argue that, at least for the range of masses probed by the BGC, there is no evidence for a population of intergalactic HI clouds without stellar components (Koribalski \etal ~2004).

Despite the focus on HI blind surveys, a large body of HI data for optically selected galaxies already exists.  If all (or nearly all) objects in the BGC mass range do indeed have optical counterparts, then a sufficiently deep HI sample of optically selected targets may be able to sample a population of objects similar to that of the BGC.  In particular, a sample that includes more sources and covers a greater volume than the BGC could allow for an investigation of the dependence of the HIMF on environment, as well as an independent check on the morphological dependence of the HIMF reported by Z03.

The first two investigations of the HIMF based on HI observations of a large sample of optically selected galaxies were Briggs \& Rao (1993) and SGH.  However, the former (based on 355 galaxies drawn from the catalog of Fisher \& Tully 1981) is limited to a very shallow volume ($cz<1000$ km/s).  The latter (based on 532 galaxies drawn from the Arecibo General Catalog) was limited to morphological types Sa-Sc, and had significant uncertainty in the estimate of the low-mass end of the HIMF, owing to the fact that the sample did not include any objects with $M_{HI} < 10^{8.5} M_{\odot}$.

In this paper we present an HIMF derived from an optically selected sample of spirals and irregulars with both optical diameter and HI flux limits.  The sample contains more than 2.7 times as many galaxies as any other used to derive an HIMF to date, and occupies a volume that greatly exceeds that of the BGC.  We do not sample as deeply in HI mass however.  We sample down to $M_{HI} = 10^{7.4} M_{\odot}$, whereas Z03 reaches just below $M_{HI} = 10^{7.0} M_{\odot}$.  Because of the large volume sampled, we also investigate the dependence of the HIMF on morphology and environment, and compare our results to those of other authors.

\section {The Data}
\subsection {Sample Selection}

Like SGH, our sample is drawn from the all-sky private database maintained by M.P.H. and R.G. known as the Arecibo General Catalog (AGC).  M.P.H. and R.G. also maintain a digital archive of $\sim$9100 HI line spectra of AGC galaxies, first described in Haynes \etal (~1999).  The majority of the galaxies in the archive are in the `Arecibo sky' ($-2^{\circ} <$ decl. $< +38^{\circ}$).  HI observational data for galaxies in that region was presented in Giovanelli \& Haynes (1985a), Giovanelli \etal ~(1986), Giovanelli \& Haynes (1989), Wegner, Haynes, \& Giovanelli (1993), Giovanelli \& Haynes (1993), Giovanelli \etal ~(1997), Haynes \etal ~(1997), and Haynes \etal ~(1999) and also includes new observations of 120 galaxies made with the Arecibo 305 m telescope of the National Astronomy and Ionosphere Center.  All of these data have been newly reprocessed, using the HI flux and width corrections described in Haynes \etal ~(1999).  This homogeneous, archival dataset is being prepared for publication (Springob, Haynes, \& Giovanelli 2004, in preparation) and access to the digital spectral archive will be provided through the U.S. National Virtual Observatory.  As in Haynes \etal ~(1999), we augment the sample of galaxies in the archive with HI fluxes from the literature as compiled in the AGC.  The total number of HI flux measurements available in this sample is 13,773.

Both methods used to derive the HIMF (described in Section 3) require an accurate characterization of the selection criteria of the galaxy sample.  We thus need to select a subset of the galaxies from the archive which is complete for a particular set of criteria.  The AGC contains all objects in the {\it Uppsala General Catalog of Galaxies} (hereafter UGC, Nilson 1973), which claims to be complete down to an apparent diameter of $1.0'$ for $decl. > -2.5^{\circ}$.  Thus, almost all AGC galaxies in the Arecibo declination range with apparent diameter greater than $1.0'$ are UGC objects.  There are a few exceptions to this, as the UGC is not {\it quite} complete close to this diameter limit.  In some instances, double galaxies listed as single in the UGC are listed singly in the AGC, and some additional objects, notably those drawn from the Catalog of Galaxies and Clusters of Galaxies (Zwicky \etal ~1960-1968) that are large enough ($a>1.0'$) are included.  The diameter completeness of the UGC is discussed in detail in Hudson \& Lynden-Bell (1991).

We therefore select a sample of 2771 galaxies (hereafter referred to as the ``complete sample'', as distinct from the subsamples discussed in Sections 4.1 and 4.2), containing all objects in the HI digital archive or the AGC with $a > 1.0$ (where $a$ is the optical apparent diameter in arcminutes), $-2^{\circ} <$ decl. $< +38^{\circ}$, galactic latitude $|b| > 15^{\circ}$, morphological type Sa-Irr, and for which log$(F_{HI})>0.6$ (where $F_{HI}$ is the corrected HI flux density integrated over the profile in units of Jy km/s).  Ellipticals, S0s, and S0as are excluded due to the poor completeness of the archive for these types (see Section 2.3).  The choice of HI flux limit is explained in Section 2.3.

\subsection {Galaxy properties}

\subsubsection {Optical properties}

Morphological types and optical diameters were derived using the same methods described in Section 2.2 of SGH.  As in SGH, we define the linear optical diameter, $D_0$, as the physical size of major diameter $a$ as seen at a radial distance $r$.  This is given by the relation

\be
(D_0 / {\rm kpc}) = 0.291(r)a
\ee

\noindent where $r$ is expressed in units of Mpc.  A Hubble constant of $H_0 = 70$ km/s Mpc$^{-1}$ is assumed here and throughout this paper.

\subsubsection {Radial distance}

Heliocentric radial velocities are derived from the HI line spectrum.  The distribution of the systemic velocities for the complete sample is shown in the upper left panel of Figure 1. It reflects strongly the principal features in the local large scale structure in this area of sky, most notably the Local, Pisces--Perseus and Coma--A1367 superclusters.  As Masters, Haynes, \& Giovanelli ~(2004) point out, however, local departures from Hubble flow are likely to be significant enough that we cannot simply use the redshift distance for each galaxy.  We thus convert the velocities into distances using the multi-attractor flow model of Tonry \etal ~(2000), modified to assume our value  of $H_0 = 70$ km/s Mpc$^{-1}$.  The model includes infall onto Virgo and the Great Attractor, but reduces to $v/H_0$ at large distances.

\begin{figure}
\figurenum{1}
\epsscale{1.0}
\plotone{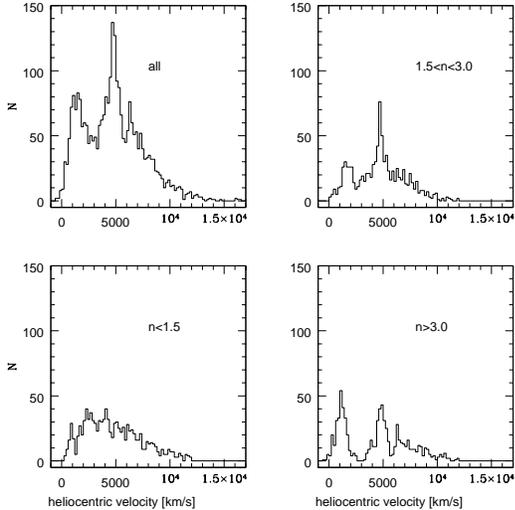}

\figcaption{Distribution of heliocentric radial velocities for all galaxies in both the complete sample and the $n<1.5$, $1.5<n<3.0$, and $n>3.0$ subsamples, in bins of width 200 km/s.  The peak velocity of the complete sample at $\sim4800$ km/s is roughly a factor of three greater than that of the BGC (see Koribalski \etal ~2004), the largest HI selected sample for which a HIMF has been derived.  The mean velocity is similar to that of the Arecibo Dual-Beam Survey used by RS (see Rosenberg \& Schneider 2000 Figure 12), but our sample provides much better statistics.  Note how the denser subsamples are ``clumpier'' than the lowest density regions.\label{FIG1}}
\end{figure}

For galaxies in clusters (or large groups), however, the peculiar velocity induced by the cluster or group is likely to make the Tonry model applied to the {\it galaxy} redshift an insufficient estimate of the distance.  Thus, where possible, we assign group or cluster membership to galaxies and use the group or cluster redshift instead.  68 of the galaxies in our sample have been identified as cluster galaxies by either Giovanelli \etal ~(1997) or Dale \etal ~(1999).  For each of these objects, we use the cluster redshift.  An additional 777 galaxies are either identified in the ``P2'' Nearby Optical Galaxies (NOG) group catalog (Giuricin \etal ~2000), or were found to be NOG group members via a ``friends-of-friends'' algorithm similar to that used by the P2 catalog.  For these objects, we use the NOG group redshift.

\subsubsection {HI Mass}

The neutral hydrogen mass of each source is given by

\be
M_{HI}/M_{\odot} = 2.36 \times 10^{5} r^2 F_{HI}
\ee

HI fluxes are corrected for beam attenuation, HI self-absorption, and pointing offsets as described in Springob, Haynes, \& Giovanelli (2004, in preparation).

\subsection {Completeness} 

The archive contains HI data for 74\% of Sa-Sb galaxies, 88\% of Sbc-Sc galaxies, 81\% of Scd-Sd galaxies, and 97\% of Sm-Irr AGC galaxies with diameter $a>1.0$ \arcmin~ in the sky region $-2^{\circ} <$ decl. $< +38^{\circ}$, $|b| > 15^{\circ}$.  In contrast, it contains a mere 14\% of E-S0a galaxies fulfilling these criteria.  These fractions include nondetections and obvious cases of HI confusion.  However, as shown in Figure 2, virtually all of the galaxies without HI data lie close to the diameter limit.  Imposing a suitable HI flux limit should improve the completeness considerably.

\begin{figure}
\figurenum{2}
\epsscale{1.0}
\plotone{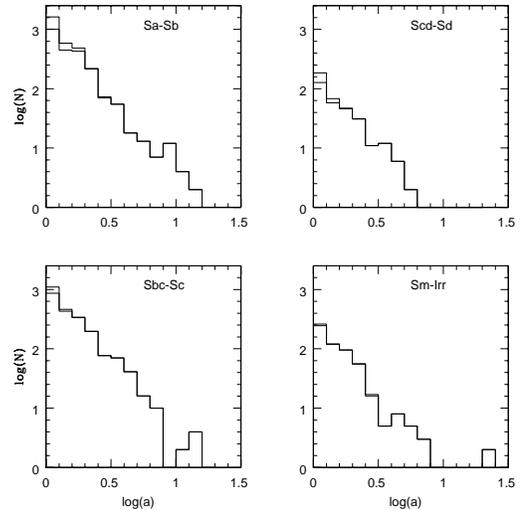}

\figcaption{Diameter completeness for different morphological subsamples.  Thin line shows distribution of apparent diameters for all galaxies with $a > 1.0$ in the AGC in the sky region $-2^{\circ} <$ decl. $< +38^{\circ}$ and $|b| > 15^{\circ}$.  Thick line shows the number of those galaxies for which there are HI parameters in the HI archive.  Bins have width 0.1 dex.  The distributions do not show any noticeable deviations from the overall diameter distribution for the UGC presented in Hudson \& Lynden-Bell 1991. \label{FIG2}}
\end{figure}

We can crudely estimate the HI flux completeness of the sample by relying on the linear relationship between log$(D_0)$ and log$(M_{HI})$ given by Haynes \& Giovanelli (1984).  This relationship is plotted for our sample in Figure 3.  A linear least-squares fit gives log$(D_0 /$kpc$) = 0.46$ log$(M_{HI}/M_{\odot}) - 3.06$, with an average scatter of 0.12.  Since the scatter is nearly Gaussian, we can estimate, for any galaxy of HI mass $M_{HI}$, the fractional number of galaxies with that mass, at that distance, which are not included in the sample because they fall below the diameter limit.  From this we can infer a  ``corrected flux distribution'' (Figure 4), the distribution of the HI fluxes that we would have if there were no diameter limit.  For an HI flux-limited sample, the logarithm of the number of galaxies per unit log$(F_{HI}$) should be proportional to -1.5 log$(F_{HI})$.  As Figure 4 shows, if one normalizes this relation to include the same number of galaxies as the actual data, the completeness drops to less than 80\% per bin below log$(F_{HI})=0.6$, so we adopt this as our HI flux limit.  In contrast, SGH actually used a {\it lower} flux limit of log$(F_{HI})=0.4$, but did not reach nearly as far down in HI mass due to their reduced sky coverage and restriction to types Sa-Sc.  Z03 used a {\it peak} flux limit, rather than an integrated flux limit as we have done, of 116 mJy.  As seen in Figure 9 of Koribalski \etal ~2004, this means that all but a handful of BGC objects have integrated fluxes of at least two or three times greater than our limit of $10^{0.6}$ Jy km/s.

\begin{figure}
\figurenum{3}
\epsscale{1.0}
\plotone{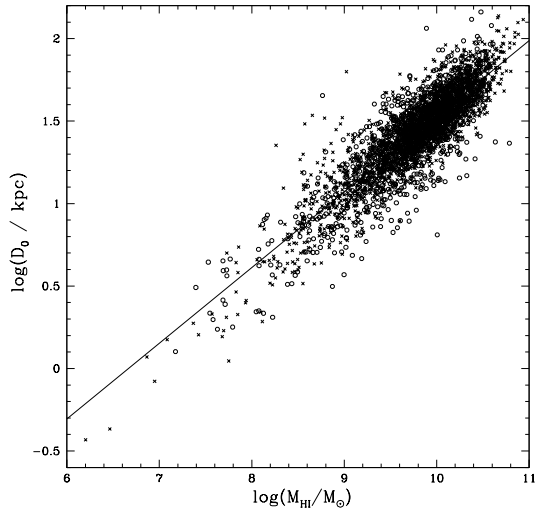}

\figcaption{Logarithm of linear optical diameter, $D_0$, vs. logarithm of HI mass, $M_{HI}$.  Crosses give the values for all galaxies in our complete sample.  Open circles are the values for the BGC for which optical diameters can be found in the AGC.  Solid line shows best-fit linear relationship between $D_0$ and $M_{HI}$ for our sample only, log$(D_0 /$ kpc) = 0.46 log$(M_{HI}/M_{\odot}) - 3.06$.\label{FIG3}}
\end{figure}

\begin{figure}
\figurenum{4}
\epsscale{1.0}
\plotone{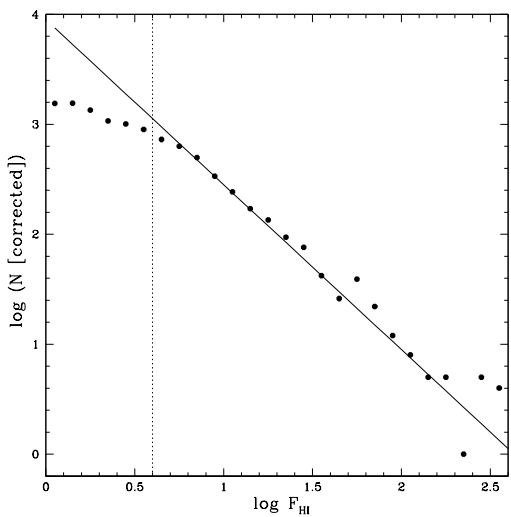}

\figcaption{Logarithmic ``corrected flux distribution''.  Filled circles show the number of galaxies per 0.1 dex bin in our sample, corrected to include galaxies that are missed due to the diameter limit.  A flux limited sample should give $d($log $N_{galaxies})/d($log $F_{HI}) \propto -1.5$ log $F_{HI} + constant$.  This solid line shows this relation, normalized to include an equal number of objects as the actual data for log $F_{HI} > 0.6$.
\label{FIG4}}
\end{figure}

The AGC includes 107 galaxies with $a > 1.0$, $-2^{\circ} <$ decl. $< +38^{\circ}$, galactic latitude $|b| > 15^{\circ}$, and type Sa-Irr for which HI observations yielded nondetections.  However, only 7 of these galaxies have an rms noise per channel greater than 5 mJy.  If we were to assume that the emission profiles for these galaxies are rectangular with amplitude 1.5 times the rms noise per channel, then a galaxy with rms noise equal to 5 mJy would require a velocity width of at least 530 km/s to be included in a flux limited sample with flux limit $10^{0.6}$ Jy km/s.  It thus seems exceedingly unlikely that the nondetections would have much impact on our result, so we exclude them from the derivation of the HIMF explained in Section 3.

There are also 55 galaxies that would otherwise meet our selection criteria, but are excluded because they are clearly confused, and it is impossible to determine how much of the HI flux belongs to the target galaxy.  Both the redshift and optical diameter distributions of these galaxies seem to match that of the complete sample, so we do not believe the exclusion of these galaxies has a significant effect on the HIMF shape.

\section {The HI Mass Function}

\subsection {Methods}

\subsubsection {$\Sigma (1/V_{max})$ Method}

The $\Sigma (1/V_{max})$ method, first developed by Schmidt (1968), has frequently been applied to the determination of HI mass functions (e.g., Zwaan \etal ~1997; RS).  Each galaxy is assigned a weight given by $1/V_{max}(x_1,x_2,...)$, where $V_{max}(x_1,x_2,...)$ is the effective search volume for a galaxy with a particular set of physical properties $x_1, x_2....$ relevant to the observational selection criteria.  The HIMF, $\phi(M_{HI})$, is defined as the space density of objects with mass $M_{HI}$.  And so the value of $\phi(M_{HI})$ for a particular mass bin is given by the sum of the weights of all galaxies in that bin.

In our case, the selection function depends on optical diameter and HI flux.  Following equations 1 and 2, we see that

\begin{eqnarray}
V_{max}(D_0,M_{HI}) = {{4 \pi f_{sky}} \over 3} \biggl(min \biggl[{ D_0 \over {\rm kpc} } {1 \over 0.291 a_{lim}}, \nonumber \\
\sqrt{{M_{HI} \over M_{\odot}} {1 \over 2.36 \times 10^{5} F_{HI lim}}} \biggr] \biggr)^3 {\rm Mpc}^3
\end{eqnarray}

\noindent where $a_{lim}$ and $F_{HI lim}$ are the optical diameter and HI flux limits ($1'$ and $10^{0.6}$ Jy km/s respectively), and $f_{sky}$ is the fraction of the sky covered by the survey (equal to 0.257 for $-2^{\circ} < decl. < 38^{\circ}$ and $|b| > 15^{\circ}$).

The primary concern about the $\Sigma (1/V_{max})$ method is the implicit assumption that sources are distributed homogeneously within the survey volume.  The survey volume for our sample contains rather significant density inhomogeneities, most notably, the Local Supercluster and the Pisces-Perseus Supercluster, so noticeable in Figure 1.  To offset this effect, modified versions of $\Sigma (1/V_{max})$ which account for the effects of large scale structure have been used by RS and Masters, Haynes, \& Giovanelli (2004).  In both of these cases, each galaxy was weighted by $1/V_{max-effective} = 1/ [n(V_{max}) V_{max})]$ rather than $1/V_{max}$, where $n(V_{max})$ is the average density within the volume $V_{max}$ normalized by the average density of the universe.

However, while RS derived $n(V_{max})$ from a magnitude-limited sample of galaxies drawn from the NASA/IPAC Extragalactic Database, Masters, Haynes, \& Giovanelli (2004) used the matter density reconstruction derived from the IRAS Point Source Catalog Redshift Survey (PSCz).  The latter reduces the likelihood that the estimate of $n(V_{max})$ will be influenced by distortions in redshift space due to galaxy peculiar velocities, since such peculiar velocities are accounted for by the density reconstruction.  We have therefore chosen to use the Masters, Haynes, \& Giovanelli (2004) approach.

The PSCz density reconstruction (presented in Branchini \etal ~1999) consists of a cartesian grid of mass densities (with grid spacings of 1.34 Mpc out to a limit of 85.7 Mpc) derived from the distribution of PSCz sources in redshift space, assuming $\beta = \Omega_{mass}^{0.6}/b = 0.5$.  At radial distance steps of 1.34 Mpc, we average all PSCz density gridpoints interior to the radial distance at the step that occupy $-2^{\circ} <$ decl. $< 38^{\circ}$ and $|b| > 15^{\circ}$.  The average density for that volume is taken to be $n(V_{max})$.  Beyond 85.7 Mpc, we assume $n(V_{max})$ = 1.

\subsubsection {2DSWML Method}

An alternative method for deriving the HIMF is the two-dimensional stepwise maximum likelihood (2DSWML) method, developed by Loveday (2000) for constructing an optical luminosity function, and first used for the HIMF by Z03.  Z03 used this method rather than $\Sigma (1/V_{max})$ because of the latter's insensitivity to the effects of large scale structure.  We believe this concern has been abrogated by our use of a modified version of $\Sigma (1/V_{max})$ that accounts for large scale structure as described in the previous subsection.  2DSWML has its own drawbacks: Unlike $\Sigma (1/V_{max})$, it is not automatically normalized and it also assumes that the shape of the mass function is the same everywhere.  The latter effect can skew the HIMF at the high-mass end to match this assumption.

Nevertheless, we have implemented the 2DSWML method as a check on our result.  Rather than binning the sample by HI mass and HI velocity width, as Z03 have done, we have binned by HI mass and linear optical diameter.  Bins are 0.25 dex wide in logarithmic mass (just as with the $\Sigma (1/V_{max})$ method) and 0.1 dex wide in logarithmic diameter.  A detailed description of the method is provided by Z03.

\section {Results}

Figure 5 shows the HIMF, as determined by both the $\Sigma (1/V_{max})$ and 2DSWML methods, for the complete sample.  The normalization for the 2DSWML points has been chosen to minimize scatter with the $\Sigma (1/V_{max})$ points.  Agreement between the two methods appears to be quite good, with the 2DSWML points falling within 1$\sigma$ of the $\Sigma (1/V_{max})$ for 12 of the 15 bins (and within 0.2$\sigma$ for 7 of those bins).  We note, however, that the high-mass end of the 2DSWML HIMF is influenced by values in lower mass bins (as mentioned in Section 3.1.2).  The two bins that show the greatest deviation from the $\Sigma (1/V_{max})$ results are in fact the two highest mass bins.  Since the earliest types tend to have very little HI (see, e.g., Roberts \& Haynes 1994), we expect that the exclusion of types earlier than Sa makes little difference in the resulting HIMF.

\begin{figure}
\figurenum{5}
\epsscale{1.0}
\plotone{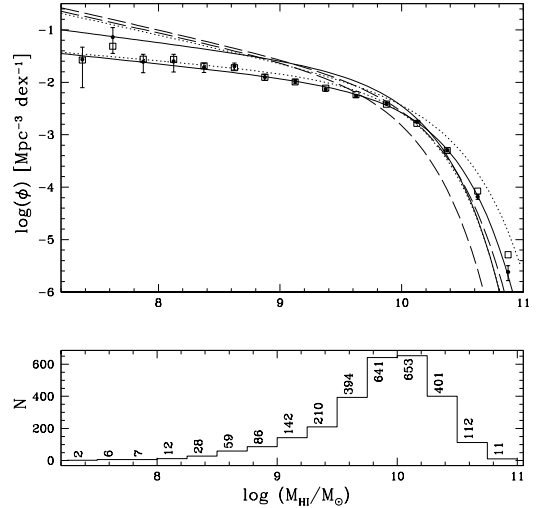}

\figcaption{{\it Top}, HIMF for the complete sample, using the $\Sigma (1/V_{max})$ method (filled circles) and the 2DSWML method (open squares).  $\Sigma (1/V_{max})$ errorbars indicate 1$\sigma$ uncertainties based on Poisson counting statistics.  The thick solid line is the best-fit Schechter function to the $\Sigma (1/V_{max})$ points: $\alpha = -1.24$, log$(M_{*}/M_{\odot})=9.99$, $\phi_{*}=3.2 \times 10^{-3}$ Mpc$^{-3}$.  Other lines show the Schechter parameters from five HI blind surveys.  (Zwaan \etal ~1997: {\it thick dotted}; RS: {\it thin dashed}; Henning \etal ~2000: {\it thick dashed}; Kilborn 2000: {\it thin dotted}; Z03: {\it thin solid}); {\it bottom}, distribution of HI masses in the full sample per 0.25 dex bin.  Comparing this distribution to Rosenberg \& Schneider (2000) Figure 16 and Z03 Figure 4, one can see that, while we have many more objects across most of the mass range, we do not go any lower in mass than those surveys, and in fact have fewer objects at the lowest masses than Z03.
\label{FIG5}}
\end{figure}

We have fit the $\Sigma (1/V_{max})$ HIMF to a Schechter function (Schechter 1976) defined by

\begin{eqnarray}
\phi(M_{HI}) = {dN \over d \; {\rm log}(M_{HI})} \nonumber \\
= {\rm ln}(10) \phi_{*} (M_{HI}/M_{*})^{\alpha + 1} {\rm exp} (-M_{HI}/M_{*})
\end{eqnarray}

\noindent Our best fit Schechter parameters are $\alpha = -1.24$, log$(M_{*}/M_{\odot})=9.99$, $\phi_{*}=3.2 \times 10^{-3}$ Mpc$^{-3}$, with a $\chi_\nu^2$ of 0.99.  Our $\Sigma (1/V_{max})$ errorbars are based on Poisson counting statistics, and ignore any errors introduced by binning.  We have experimented with varying the bin sizes across the range of $M_{HI}$ (so that the bins are more finely spaced for higher values of $M_{HI}$).  This does yield a modest decrease in ${\chi_\nu}^2$, but no significant change in the values of the Schechter parameters, so we stick with the fixed bin widths for the sake of simplicity.

Best fit values for $\alpha$ and $M_{*}$, along with error contours (based on $\chi^2$ goodness of fit), are shown in Figure 6.  For comparison we plot the results for the five HI selected samples listed in Table 3 of Z03 (i.e., the results of Zwaan \etal ~1997, Kilborn 2000, Henning \etal ~2000, and RS, as well as Z03 itself).  The agreement between our Schechter parameters and those of HI selected samples is no worse than the agreement of the latter's Schechter parameters with one another.  As Masters, Haynes, \& Giovanelli (2004) point out, however, the slope of the Z03 HIMF is actually steeper when one takes into account the effects of peculiar velocities.  The authors estimate that a flow model-corrected Z03 HIMF would actually have a low mass slope of -1.4, which would put our result on the ``shallow'' end of HIMF results along with Zwaan \etal ~(1997).  In fact, even if we accept the Z03 HIMF as published, one must account for the authors' high value for the normalization constant $\phi_{*}$ in making any comparison of the HIMFs at the low mass end.  In Figure 5, we have overplotted the HIMFs of all five HI blind samples using the Schechter parameters as published.  At masses of $\sim 10^8 M_{\odot}$, both our HIMF and that of Zwaan \etal ~(1997) are lower than each of the others by a factor of a few.  Another way to compare the results is by comparing the total HI mass density, derived by integrating the Schechter function over all HI masses.  Using $\rho_{HI}=\phi_{*}\Gamma(2+\alpha)M_{*}$, we get $\rho_{HI}=3.8 \times 10^7 M_{\odot}$/Mpc$^{-3}$.  Converting this to $\Omega_{HI}$ gives $\Omega_{HI} = 2.7\times 10^{-4}$, 39\% lower than the value reported by Z03.

\begin{figure}
\figurenum{6}
\epsscale{1.0}
\plotone{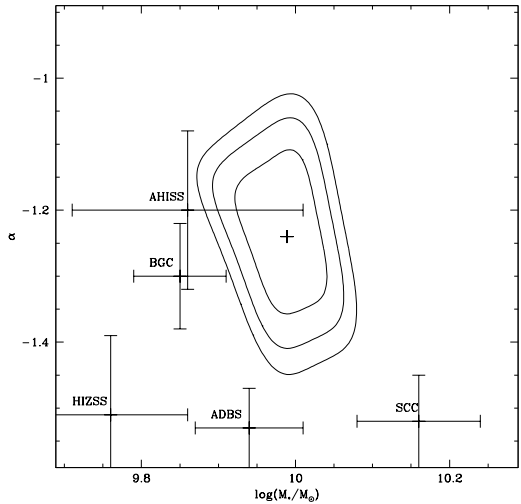}

\figcaption{The $\chi^2$ contours of the Schechter fit to the HIMF for the complete sample.  Contours represent 1, 2, and 3$\sigma$.  Also shown are Schechter parameters from five HI blind surveys: the Arecibo HI Strip Survey (AHISS, Zwaan \etal ~1997), the Arecibo Dual-Beam Survey (ADBS, RS), the HIPASS HIZSS (Henning \etal ~2000), the HIPASS SCC (Kilborn 2000), and the HIPASS Bright Galaxy Catalog (BGC, Z03).  ADBS errorbars are estimated from the contours in RS Figure 3.  AHISS errorbars are from Zwaan (2004, private communication).
\label{FIG6}}
\end{figure}

There are several possible explanations for this discrepancy.  It may be that galaxies with small physical diameters, not included in our complete sample because of our diameter limit, are large contributors to the low mass end of the HIMF.  In Figure 3, we have overplotted log$(D_0)$ vs. log$(M_{HI})$ for all BGC galaxies for which there are optical diameters listed in the AGC.  As with our own data, the values of $D_0$ and $M_{HI}$ are calculated using the Tonry model-corrected distance.  It does appear that across most of the $M_{HI}$ distribution, the BGC includes a population of galaxies with smaller diameters than those included in our complete sample.  However, this is least apparent for log$(M_{HI}/M_{\odot}) < 8.5$, where the divergence between the HIMFs is greatest.

We also note that Briggs \& Rao (1993),  the only other work to estimate the HIMF based on a large optically selected sample including the range of morphological types we include here, finds a low mass slope of $\alpha \sim -1.25$, nearly identical to our value.  This suggests that the discrepancy may be due to some hidden systematic bias in the selection of optical samples.  HI selected samples may be detecting faint low surface brightness galaxies that are missed by optical surveys.  This could include galaxies that meet our diameter limit, but are nonetheless not included in our sample because they are too faint to be optically detected.  Yet we find that, of the 758 BGC galaxies with $|b| > 15^{\circ}$, only 39 do not have optical counterparts with diameter estimates listed in the AGC.  And only 3 of those have log$(M_{HI}/M_{\odot}) < 8.5$.

Another possibile explanation for the disagreement between our estimate of the low mass end of the HIMF and those of most HI selected samples is that the discrepancy is a consequence of the fact that the lowest mass galaxies in each of these samples are confined to a very local volume.  In both the BGC and our complete sample, all galaxies with log$(M_{HI}/M_{\odot}) < 8.5$ are found within $v_{helio} < 1200$ km/s.  There could be systematic distance errors due to defects in the flow model, or there could be systematic environmental effects within that local volume.  We note that the only HIMF derived from an HI blind survey which matches our result at low masses is Zwaan \etal ~(1997), which, like our sample and unlike HIPASS, includes nearby galaxies in the direction of Virgo.

It should be noted that the Zwaan \etal ~(1997) sample, like the majority
of ours, was observed with the 305m Arecibo telescope, which has a much 
narrower beam (3.3\arcmin) than the Parkes 64m radio telescope used for 
HIPASS (15.5\arcmin). Additionally, with few exceptions, interacting and 
highly disturbed systems are not included in our sample. We thus do not believe that confusion has any meaningful impact on the HIMF shape we have derived
for our sample, but it may have some impact on the HIMFs derived from
HIPASS.  It is not clear whether confusion would systematically effect low
or high mass objects however.

Given both the distance uncertainties and our poor understanding of the environmental dependence of the HIMF (which we discuss in Sections 4.2 and 5), we cannot resolve the question of why our HIMF diverges from that of most HI blind surveys at the low mass end.  These effects can only be disentangled with the aid of a deeper sample that includes low mass galaxies at larger distances.

\subsection {Morphological Dependence}

We have computed the HIMF for four subsets of the complete sample, divided by morphological type: Sa-Sb, Sbc-Sc, Scd-Sd, and Sm-Irr.  The results are plotted in Figure 7, and best-fit Schechter parameters are given in Table 1.  Figure 8 shows the 1 $\sigma$ contour plots, along with the corresponding results from Z03 and SGH.  (Our morphological types were derived in the same manner as in SGH, i.e., primarily using UGC types.  Z03, on the other hand, uses Lyon/Meudon Extragalactic Database types.)  Our results are broadly consistent with those authors for most morphological types.  Types Sa-Sc occupy a parameter space that is clearly distinct from types Scd and later: The former grouping has higher values of both $M_{*}$ and $\alpha$ than the latter.

\begin{figure}
\figurenum{7}
\epsscale{1.0}
\plotone{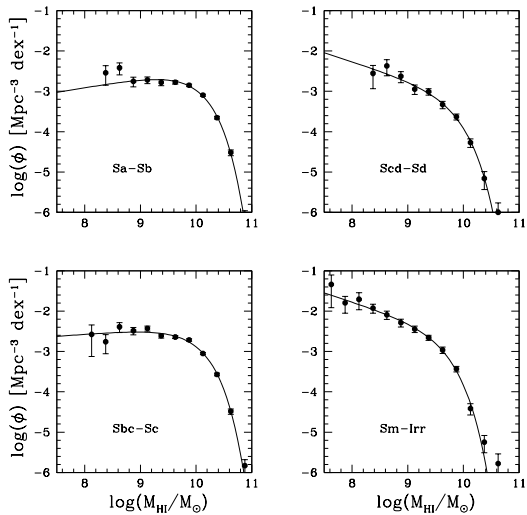}

\figcaption{HIMFs for the each of the morphological subsamples.  Solid lines show best-fit Schechter functions.\label{FIG7}}
\end{figure}

\begin{deluxetable}{lccccc}
\tablewidth{0pt}
\tablenum{1}
\tablecaption{Schechter Parameters\label{Tab1}}
\tablehead{
\colhead{subsample}   & \colhead{$N_{galaxies}$}     & \colhead{$\alpha$}   & \colhead{log$(M_{*}/M_{\odot})$}  & \colhead{$\phi_{*} [$Mpc$^{-3}]$}    & \colhead{${\chi_\nu}^2$}                     
}
\startdata
all  & 2771  &  -1.24 & 9.99 &  $3.2 \times 10^{-3}$ &  0.99\nl
Sa-Sb  & 1003  &   -0.77 & 9.91 &  $1.5 \times 10^{-3}$ &  0.97\nl
Sbc-Sc  & 1261  &   -0.90 & 9.92 &  $1.8 \times 10^{-3}$ &  1.24\nl
Scd-Sd  & 176  &   -1.44 & 9.75 &  $4.0 \times 10^{-4}$ &  0.56\nl
Sm-Irr  & 331  &   -1.44 & 9.57 &  $1.5 \times 10^{-3}$ &  0.64\nl
$n<1.5$  & 1036  &   -1.38 & 10.07 &  $8.7 \times 10^{-4}$ &  0.35\nl
$1.5<n<3.0$  & 918  &   -1.13 & 9.92 &  $8.2 \times 10^{-3}$ &  0.71\nl
$n>3.0$  & 751  &   -1.24 & 9.95 &  $2.1 \times 10^{-2}$ &  0.65\nl

\enddata
\end{deluxetable}

\begin{figure}
\figurenum{8}
\epsscale{1.0}
\plotone{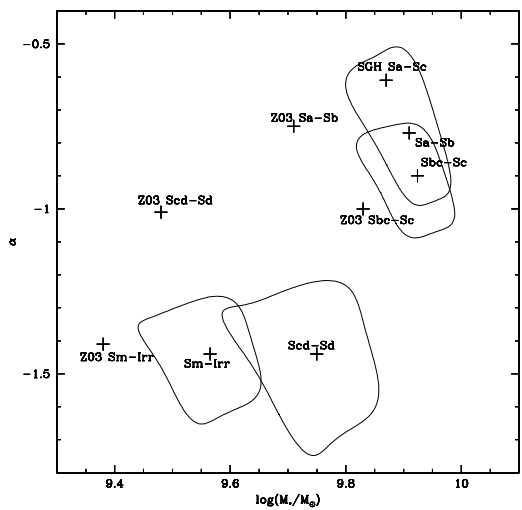}

\figcaption{The $\chi^2$ contours of the Schechter fit to the HIMF for morphological subsamples Sa-Sb, Sbc-Sc, Scd-Sd, and Sm-Irr.  Contours represent 1 $\sigma$ uncertainties. Corresponding parameters from Z03 and SGH are also shown.\label{FIG8}}
\end{figure}

Z03 makes no estimate of the uncertainty on their best fit Schechter parameters for their morphological subsamples.  However, it appears that if their uncertainties are similar in magnitude to ours, then the 1 $\sigma$ uncertainties would overlap or nearly overlap for all types with the possible exception of Scd-Sd.  The discrepancy for that case could be due to differences in the morphological type assignment criteria.

The higher values of both $M_{*}$ and $\alpha$ for Sa-Sc galaxies than for later types is not unexpected, given the decrease in the average HI mass for later types relative to Sa-Sc shown in Table 1 of Roberts \& Haynes (1994).

\subsection {Environmental Dependence}

We have also computed the HIMF for three subsets of the complete sample, divided by local matter density as determined by PSCz ($n<1.5$, $1.5<n<3.0$, and $n>3.0$).  For this task, the version of the PSCz density reconstruction referred to in Section 3.1.1 is insufficient, due to its 85.7 Mpc limit.  We therefore use an alternate version of the density reconstruction, with grid spacings of 2.68 Mpc out to a limit of 171.4 Mpc.

In order to compute the HIMF for each of these subsets using the $\Sigma (1/V_{max})$ method, we must account for the fact that, not only does the space density of objects vary with distance due to large scale structure, but the fractional volume of space occupied by regions of a particular density does as well (see Figure 1).  We thus compute the HIMF for each of our three subsamples using the effective $V_{max}$

\begin{eqnarray}
{V_{max-effective}} = min\biggl[(171.4 ~{\rm Mpc})^3, \nonumber \\
{f_{int}(V_{max}) \over n_{int}(V_{max})}n(V_{max})V_{max}\biggr]
\end{eqnarray}

\noindent where $V_{max}$ is given by Equation 3, $f_{int}(V_{max})$ is the fraction of the volume $V_{max}$ occupied by regions of the density range for each given subsample, and $n_{int}(V_{max})$ is the average density within those regions.  $n(V_{max})$ now represents the average density within regions of the density range for each given subsample, {\it not} within the entire volume $V_{max}$.  Since PSCz only extends to 171.4 Mpc, all galaxies with distances greater than 171.4 Mpc are discarded, and $V_{max-effective}$ is capped at that distance.

The resulting HIMFs are shown in Figure 9, with corresponding $\chi^2$ contour plots in Figure 10.  Best-fit Schechter parameters can be found in Table 1.  We find that the lowest density sample has a steeper low-mass slope and a higher value of $M_{*}$ than the two higher density samples.  This trend is consistent with the results of  Briggs \& Rao (1993), RS, and Davies \etal ~(2004), all of whom found evidence for a flatter HIMF in Virgo than in the field, and Verheijen \etal ~(2001), who found a flatter HIMF in Ursa Major than in the field.  However, as seen by the overlapping 1 $\sigma$ error contours, the uncertainties are large enough that the data are consistent with the Schechter parameters being identical for all three regions.  The data are insufficient to show conclusively that the HIMF varies with environment.

\begin{figure}
\figurenum{9}
\epsscale{1.0}
\plotone{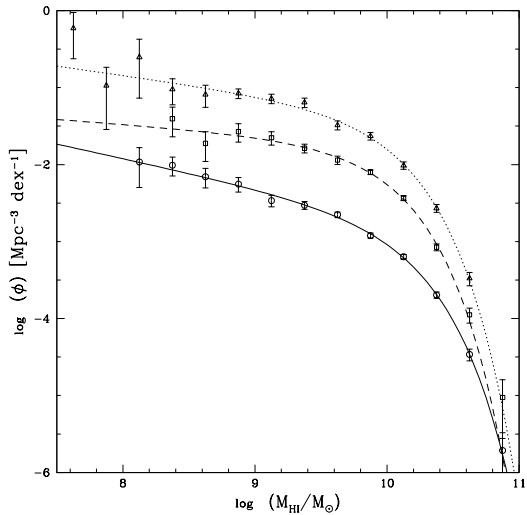}

\figcaption{HIMFs for the $n<1.5$ (open circles), $1.5<n<3.0$ (open squares), and $n>3.0$ (open triangles) samples.  Solid, long-dashed, and short-dashed lines show best-fit Schechter functions for each of the three samples respectively.\label{FIG9}}
\end{figure}

\begin{figure}
\figurenum{10}
\epsscale{1.0}
\plotone{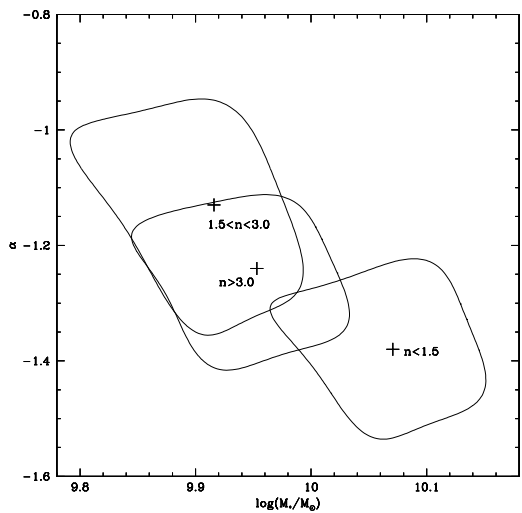}

\figcaption{The $\chi^2$ contours of the Schechter fit to the HIMF for the $n<1.5$, $1.5<n<3.0$, and $n>3.0$ samples.  Contours represent 1 $\sigma$ uncertainties. \label{FIG10}}
\end{figure}

We also note that, because the density field is derived from a flux limited catalog (and smoothed with a Gaussian filter of {\it constant radius} 4.57 Mpc) in which the average galaxy-galaxy separation increases with distance, the shot noise of the density field {\it increases} with distance.  Thus, for example, the set of all gridpoints with observed $n>3.0$ at large distances will contain more contamination from less dense regions than the corresponding set of gridpoints with observed $n>3.0$ at nearby distances.  This effect {\it may} have some influence on our results, though it appears not to create any deviation of the HIMF shape from a Schechter function.  As seen in Table 1, $\chi_\nu^2$ is actually {\it lower} for all three of the density-selected subsets than it is for the complete sample.

Finally, we check whether the apparent change in the slope of the low mass end of the HIMF is a consequence of the correlation between $\alpha$ and $M_{*}$.  If $M_{*}$ varies with environment, then it may result in a change in $\alpha$ that does not represent a true change in the low mass slope.  We have computed the least-squares best fit line for all points with log($M_{HI}/M_{\odot}$) $< 9.5$ in the log($\phi$)-log($M_{HI}$) plots of Figure 9 for each of the three density regimes.  We find slopes of -0.51, -0.32, and -0.31 for the $n<1.5$, $1.5<n<3.0$, and $n>3.0$ regions respectively, confirming that the steeper slope in the $n<1.5$ regions is not an artifact of the fitting parameters.

\section {Discussion}

Studies of the HI content of galaxies in clusters have established that a significant fraction of cluster galaxies are deficient in HI gas relative to the field (e.g., Giovanelli \& Haynes 1985b).  The most commonly suggested mechanisms invoked to explain the removal of gas from cluster galaxies involve one form or another of interaction between the HI gas in galaxies and the intercluster medium (ICM).  These mechanisms include ram-pressure stripping (Gunn \& Gott 1972), thermal evaporative stripping (Cowie \& Songaila 1977), and turbulent viscous stripping (Nulsen 1982).  Valluri \& Jog (1991) showed that all of these mechanisms should be more pronounced for galaxies with small optical sizes.  Their data, however, showed that galaxies with smaller optical sizes tend to be {\it less} deficient in HI than galaxies with larger optical sizes.  The authors used this result to rule out any mechanism for gas removal that involves galaxy-ICM interactions.  The only mechanism investigated by Valluri \& Jog (1991) that the authors concluded might be consistent with the diameter-dependent deficiency shown by their data was galaxy-galaxy tidal encounters in subclumps within the cluster.  This interpretation was criticized by Solanes \etal ~(2001), who noted that the deficiency parameter used by Valluri \& Jog (1991) had a residual dependence on the optical diameter.  The data presented by Solanes \etal ~(2001) suggested that in fact, there is no apparent dependence of HI deficiency on optical diameter.

If HI deficiency in clusters {\it does} depend on optical size, then the dependence should manifest itself in the cluster HIMF.  {\it If} the population of galaxies in clusters was otherwise identical to the population of galaxies in the field, then a constant deficiency for all cluster galaxies would mean a cluster HIMF with the same shape as the field HIMF, but shifted to lower HI masses.

Of course, not all cluster galaxies have the same HI deficiency, so we require a somewhat more sophisticated model.  If we define $\phi_{cl}(M_{HI})$ as the cluster HIMF {\it that would exist if none of the galaxies were deficient}, then we can express $\phi'_{cl}(M_{HI})$, the cluster HIMF that {\it includes} HI deficient galaxies, as

\be
\phi'_{cl}(M_{HI}) = f_0 (M_{HI}) \phi_{cl}(M_{HI}) + \sum_{i} f_i \phi_{cl}(M_{HI}+\Delta_{i})
\ee

\noindent where $f_0(M_{HI})$ is the fractional number of galaxies with HI mass $M_{HI}$ that are {\it not} HI deficient, $f_i$ is the fractional number of galaxies with expected HI mass $M_{HI}+\Delta_{i}$ that are deficient by a factor $\Delta_{i}$, and the summation is over all expected HI masses greater than $M_{HI}$.  We define the deficiency DEF as in Haynes \& Giovanelli (1984), the logarithm of the ratio of the expected HI mass for a particular linear optical diameter and morphological type to the {\it observed} HI mass for that galaxy. At the low mass end of the HIMF, where we can approximate log($\phi_{cl}(M_{HI})$) $\simeq (\alpha +1)$log($M_{HI}$)$+constant$, (6) becomes

\begin{eqnarray}
\phi'_{cl}(M_{HI}) = \biggl(f_0 (M_{HI}) + \sum_{i} f_i 10^{(\alpha+1)\Delta_{i}} \biggr) \nonumber \\
\times 10^{(\alpha+1)log(M_{HI})+constant}
\end{eqnarray}

If $f_0$ and $f_{i}$ are completely independent of linear optical diameter (that is, independent of the {\it expected} HI mass), then deficiency will not affect the low mass slope.  If, on the other hand, either the fractional number of HI deficient galaxies or the amount of HI deficiency per galaxy increases (decreases) with diameter, then the low mass slope will get steeper (shallower).

None of this analysis necessarily applies to a realistic comparison of the cluster and field HIMF.  Morphological segregation in high density regions means that the field HIMF cannot be represented by $\phi_{cl}(M_{HI})$.  As shown in Section 4.1, the HIMF for early type spirals has a much flatter low mass slope than the HIMF for later types.  So a flatter HIMF in high density regions could be a natural consequence of early type spirals being more abundant than later types in these regions.

Among the three levels of PSCz density that we study here, however, there is essentially no morphological segregation.  34\% of the $n<1.5$ galaxies are Sa-Sb, 45\% are Sbc-Sc, 6\% are Scd-Sd, and 15\% are Sm-Irr.  For the $1.5<n<3.0$ and $n>3.0$, the breakdowns are 36\%/46\%/7\%/12\% and 37\%/44\%/7\%/12\% respectively.  We have constructed synthetic HIMFs by adding linear combinations of the HIMFs for each morphological type, described in Section 4.1, in the ratios for which they appear in each of these three density regions.  We have then computed best-fit Schechter functions for each of these HIMFs using the same errorbars on each bin shown in Figure 9.  The resulting low mass slopes are $\alpha =$ -1.27, -1.21, and -1.25 for the $n<1.5$, $1.5<n<3.0$, and $n>3.0$ regions respectively, a much smaller deviation than is actually observed.  So we conclude that {\it if} the marginally flatter HIMF in the two higher density regimes is a result of increased HI deficiency relative to the low density regime, then we believe this would be an indication of increasing deficiency towards smaller optical diameters at the low mass end of the HIMF.

As a quantitative measure of deficiency, we use the expected $D_0-M_{HI}$ relations from SGH for types Sa-Sc and the relations from Haynes \& Giovanelli (1984) for types Scd-Irr.  The resulting distribution of deficiency as a function of optical diameter is shown for all three density subsamples in Figure 11.  The average deficiency for each subsample is slightly negative ($\sim -0.2$ for all three subsamples).  This is not surprising, given that HI deficient galaxies are more likely than non-deficient galaxies to fall below our HI flux limit and not make it into the sample.  We can see from these plots that the fractional number of HI deficient galaxies (DEF$>0.3$, as defined by Solanes \etal ~2001) in the $n<1.5$ and $1.5<n<3.0$ samples is relatively small (2\% in both cases).  In the $n>3.0$ sample, the number is somewhat larger (8\%), but the only apparent trend of HI deficiency is a somewhat increased number of deficient galaxies towards intermediate diameters.  Of course, our selection criteria discriminate against HI deficient galaxies with small diameters, so there could very well be a large population of deficient galaxies with small diameters that is being missed by our sample.

\begin{figure}
\figurenum{11}
\epsscale{1.0}
\plotone{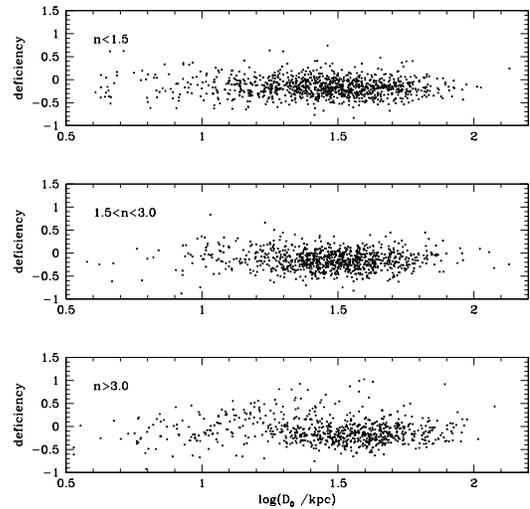}

\figcaption{Deficiency vs. linear optical diameter for the $n<1.5$, $1.5<n<3.0$, and $n>3.0$ samples. \label{FIG11}}
\end{figure}

To answer this question, we examine the sample of {\it all} AGC galaxies of types Sa-Sc with optical apparent diameter greater than 1 arcminute that can be found in the sky region covered by our complete sample.  For each of the three density regions, we compute the fraction of these galaxies with HI fluxes greater than our HI flux limit.  In Figure 12, we show the ratio of these ``HI detection fractions'' between the different density regions, divided into bins of physical linear diameter.  While the uncertainty in the ratio for each bin is significant, there is a clear trend in that the fraction of HI-detected galaxies drops towards smaller diameters more sharply in both the intermediate and high density regions than in the lowest density regions.  Since these plots are restricted to Sa-Sc and, as shown in Section 4.1, the Sa-Sb and Sbc-Sc subsamples have nearly identical HIMFs, it is unlikely that this is due to any morphological segregation effects.  In fact, this may explain why we observe so little morphological segregation among our three environmental subsamples.  As noted in Solanes \etal ~(2001), HI deficiency is preferentially found in galaxies with earlier morphological types.  While they may be more numerous in high density regions, they are also harder to detect in HI than their counterparts in low density regions.

\begin{figure}
\figurenum{12}
\epsscale{1.0}
\plotone{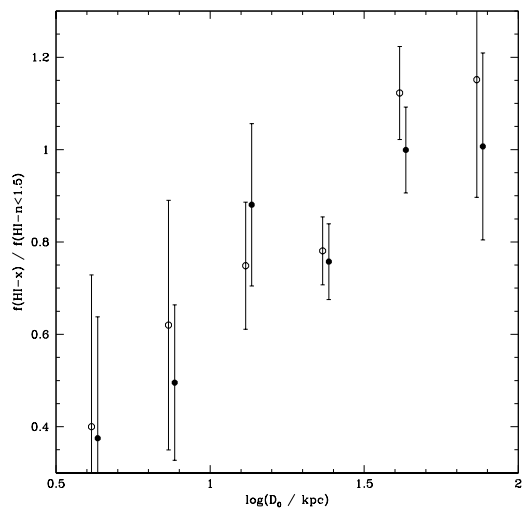}

\figcaption{Ratio of the fraction of Sa-Sc galaxies larger than 1 arcminute in optical angular diameter that meet our HI flux limit in higher density regions to that same fraction in the $n<1.5$ regions, plotted against physical optical diameter.  The filled circles are for the $n>3.0$ sample, and open circles are for the $1.5<n<3.0$.  While the errorbars are large, there is a clear trend towards lower HI detectability in smaller galaxies for the high and intermediate density regions than there is for the lowest density regions.\label{FIG12}}
\end{figure}

We thus conclude that, while the environmental dependence of the HIMF is only marginally statistically significant, the evidence argues against morphological segregation as an adequate explanation of this dependence.  It may be that the environmental dependence is caused by increased deficiency in higher density regions towards smaller linear optical diameters.  If this is the case, then it argues in favor of gas removal mechanisms that preferentially affect smaller galaxies.

We caution, however, that this conclusion depends on the premise that galaxies of a particular morphological type would have the same HIMF shape within high and low density regions were we to exclude the effects of deficiency.  Croton \etal ~(2004) shows that the optical luminosity function of the 2dF Galaxy Redshift Survey varies with environment even when one excludes the effects of morphological segregation.  There may be similar variation {\it within} the morphological subsamples of our data set that occur independently of the gas stripping that one associates with deficiency.  The statistics of our sample are insufficient to carry this line of inquiry further by segregating by morphological type more finely or segregating by morphological type and environment simultaneously.

\section {Conclusions}

We have extracted an optical diameter and HI flux-limited sample of 2771 galaxies from our HI archive to measure the HIMF of the local universe.  The best-fit Schechter parameters for this sample are $\alpha = -1.24$, log$(M_{*}/M_{\odot})=9.99$, $\phi_{*}=3.2 \times 10^{-3}$ Mpc$^{-3}$.  While the values of those parameters are as consistent with the Schechter parameters derived from HI blind survey samples as the parameters from these samples are with one another, our estimate of the number of low mass objects is lower than the number predicted by most of these other surveys by a factor of a few.  Because the low mass objects are all in the very nearby universe where distance uncertainties are greatest and environmental differences in different regions of the sky are likely to be most pronounced, it is unclear to what extent this discrepancy is caused by environmental effects, systematic errors in distance estimates, poor sampling of galaxies with small linear diameters, or the exclusion of galaxies that should meet our optical diameter limit, but are missed by optical catalogs because of low surface brightness.

Additionally, we have derived HIMFs for four morphological subsets of this sample.  The HIMF appears to have a strong morphological dependence, with earlier types having flatter low-mass ends and higher values of $M_{*}$ than later types, a result which matches that of Z03.  Our results for types Sa-Sc also appear to be roughly consistent with SGH, who found a declining low mass slope for those morphological types, despite that sample's poor sampling of the low mass end of the HIMF.

We have also performed the first comparison of the environmental dependence of the HIMF that compares all higher density regions in the survey volume to all lower density regions rather than restricting the comparison to a particular cluster or void vs. the field.  Our subsamples have been divided by local matter density as determined by the PSCz density reconstruction.  The two subsamples comprised of galaxies in higher density regions have HIMFs with flatter low-mass ends and {\it lower} values of $M_{*}$ than the subsample of galaxies in the lowest-density environments.  We have argued that the flattening of the HIMF in the two higher density regimes is unlikely to be caused by morphological segregation, and that {\it if} it is a result of increased HI deficiency relative to the low density regime, then this would be an indication of increasing deficiency towards smaller linear optical diameters at the low mass end of the HIMF, a result consistent with HI deficiency in clusters being caused by galaxy-ICM interactions.  We cannot, however, rule out the possibility that the environmental dependence is caused by variation within the morphological subsamples independent of the gas stripping in high density regions commonly associated with deficiency.  The caveat involved in any conclusions that we may draw from the differences in HIMF shape among these three subsamples is that the differences are not particularly large, and in fact the three HIMFs are marginally consistent at the $1 \sigma$ level.

The best way to resolve the question of how the HIMF depends on environment is with better statistics.  If we had a sample with many more galaxies, we could shrink the error contours in Figure 10, and we could determine which galaxies are responsible for the variation in HIMF shape by segregating by morphological type and environment at the same time, just as Croton \etal ~(2004) does with the optical luminosity function for the 2dF Survey.  In this regard, upcoming HI surveys with the Arecibo L-band Feed Array may be able to resolve the issue.

Better statistics {\it may} also be sufficient to resolve the discrepancy between our estimate of the low mass end of the HIMF and the values derived from most blind HI surveys.  With a sufficiently large HI-selected sample, one could compute the HIMF both including and excluding galaxies that were not a priori optically identified.  It could then be determined whether our estimate of the low mass end of the HIMF is depressed by the exclusion of low surface brightness galaxies not included in optical catalogs.  Likewise, better statistics would help in that they would improve our understanding of the environmental dependence, which could allow us to determine whether the low mass end of the HIMF is affected by local environmental effects.  If the discrepancy in the low mass end of the HIMF is a result of errors in the flow model, however, then a larger sample is insufficient to resolve the issue.  We would require either a better flow model or we would need a sample of low mass galaxies that goes deeper in redshift.

\vskip 0.3in

We wish to thank Enzo Branchini for providing a copy of the PSCz density field, Christian Marinoni for providing a copy of the NOG group catalog, and Karen Masters for developing software that was used for applying the flow models and fitting the Schechter functions.  This work has been partially supported by NSF grants AST-0307396, AST-0307661, the NRAO/GBT 03B-007 Graduate Student Support Grant, and the NASA New York State Space Grant.

\vfill
\end{document}